\documentclass[journal,letterpaper]{IEEEtran_nssmic}

\usepackage{cite}      

\usepackage{graphicx}  

\def\micron{{$\mu$m}}
\def\degC{{${}^\circ$C}}
\def\degree{{${}^\circ$}}
\def\cmsq{{cm${}^2$}}
\def\mmsq{{mm${}^2$}}
\def\flux{counts/s/cm${}^2$/keV}

\begin{document}

\renewcommand{\thefootnote}{\fnsymbol{footnote}}

\onecolumn

\begin{flushright}
{\small
SLAC--PUB--11828\\
April, 2006\\}
\end{flushright}

\vspace{.8cm}

\begin{center}
{\bf\large   
Design and Performance of the Soft Gamma-ray Detector
for the NeXT mission
\footnote{This work has been carried out under support of U.S. Department of Energy, contract DE-AC02-76SF00515,
Grant-in-Aid by Ministry of Education, Culture, Sports, Science and Technology of Japan (12554006, 13304014),
and ``Ground-based Research Announcement for Space Utilization'' promoted by Japan Space Forum.}}

\vspace{1cm}

H. Tajima${}^1$, T. Kamae${}^1$, G. Madejski${}^1$,
T.~Mitani${}^{2,3}$, K.~Nakazawa${}^{2}$, T.~Tanaka${}^{2,3}$, T.~Takahashi${}^{2,3}$, S.~Watanabe${}^{2}$,
Y.~Fukazawa${}^4$,
T.~Ikagawa${}^5$, J.~Kataoka${}^5$,
M.~Kokubun${}^3$, K.~Makishima${}^3$,
Y.~Terada${}^6$,
M.~Nomachi${}^7$ and
M.~Tashiro${}^8$%
\medskip

${}^1$Stanford Linear Accelerator Center, Stanford University,
Stanford, CA  94309\\
${}^2$Institute of Space and Astronautical Science, 
Sagamihara, Kanagawa 229-8510, Japan\\
${}^3$Department of Physics, University of Tokyo, 
Bunkyo-ku, Tokyo 113-0033, Japan\\
${}^4$Department of Physics, Hiroshima University, 
Higashi-Hiroshima 739-8526, Japan\\
${}^5$Department of Physics, Tokyo Institute if Technology,
Meguro-ku, Tokyo, 152-8551, Japan\\
${}^6$RIKEN, Wako, Saitama 351-0198, Japan\\
${}^7$Laboratory of Nuclear Studies, Osaka University, Toyonaka, Osaka 560-0043, Japan\\
${}^8$Department of Physics, Saitama University,
255 Shimo-Ohkubo, Saitama,  338-8570 Saitama, Japan

\end{center}

\vfill

\begin{center}
{\bf\large   
Abstract }
\end{center}

\begin{quote}
The Soft Gamma-ray Detector (SGD) on board the NeXT (Japanese future high energy astrophysics mission) is a Compton telescope with narrow field of view (FOV), which utilizes Compton kinematics to enhance its background rejection capabilities. It is realized as a hybrid semiconductor gamma-ray detector which consists of  silicon and CdTe (cadmium telluride) detectors. It can detect photons in a wide energy band (0.05--1~MeV) at a background level of $5\times 10^{-7}$~\flux; the silicon layers are required to improve the performance at a lower energy band ($<$0.3 MeV). Excellent energy resolution is the key feature of the SGD, allowing it to achieve both high angular resolution and good background rejection capability. 
An additional capability of the SGD, its ability to measure gamma-ray polarization, opens up a new window to study properties of astronomical objects. We will present the development of key technologies to realize the SGD: high quality CdTe, low noise front-end ASIC and bump bonding technology. Energy resolutions of 1.7 keV (FWHM) for CdTe pixel detectors and 1.1 keV for Si strip detectors have been measured. 
We also present the validation of Monte Carlo simulation used to evaluate the performance of the SGD.
\end{quote}

\bigskip


\vfill

\begin{center} 
{\it Contributed to} 
{\it IEEE Nuclear Science Symposium, Roma, Italia, October 16--October 22, 2004} \\
{\it Published as} 
{\it IEEE Transactions on Nuclear Science, Volume 52, Issue 6, Part 2, Dec. 2005 Pages:2749-2757  }\\



\end{center}

\twocolumn

\clearpage
\section{Introduction}
\label{sect:intro}  
%
%
\PARstart{T}{he} hard X-ray and gamma-ray bands have long been recognized as important windows for exploring the energetic universe. 
It is in these energy bands that non-thermal emission, primarily due to accelerated high energy particles, becomes dominant. 
However, by comparison with the soft X-ray band, where the spectacular data from the XMM-Newton and Chandra satellites are revolutionizing our understanding of the high-energy universe, the sensitivities of hard X-ray missions flown so far, or currently under construction, have not dramatically improved over the last decade. Clearly, the scope of discovery expected with much improved sensitivity for both point and extended sources is enormous.

The energy band between 0.1~MeV and 100~MeV is poorly explored 
due to difficulties associated with the detection of such photons.
The Compton telescope COMPTEL~\cite{COMPTEL93} on board CGRO (Compton Gamma-Ray 
Observatory) demonstrated that a gamma-ray instrument based on the Compton 
scattering is useful for the detection of gamma-rays in this energy band.
COMPTEL provided us with rich information on a variety of gamma-ray emitting 
objects either in continuum or line emission. 
The continuum sources include spin-down pulsars, stellar black-hole candidates,
supernova remnants, interstellar clouds, active galactic nuclei (AGN), 
gamma-ray bursts (GRB) and solar flares. 
Detection has also been made of the nuclear gamma-ray lines from 
${}^{26}$Al (1.809~MeV), ${}^{44}$Ti (1.157~MeV), and 
${}^{56}$Co (0.847 and 1.238~MeV). 

Although COMPTEL performed very well as the first Compton telescope 
in space for MeV gamma-ray astrophysics, it suffered severely 
from large background, poor angular resolution, and complicated image 
decoding~\cite{Knodlseder96}.
In 1987, T. Kamae {\it et al}. proposed a new Compton telescope based 
on a stack of silicon strip detectors (SSD)~\cite{Kamae87,Kamae88}.
This technology presents very attractive possibilities to overcome 
the weaknesses of COMPTEL as described later in this document.
This idea of using silicon strip detectors stimulated new proposals 
for the next generation Compton telescope~\cite{Takahashi02-NeXT,Takahashi03-SGD,MEGA,Milne,TIGRE03,NCT04}.

Recently, a new semiconductor detector based on cadmium telluride (CdTe) 
emerged as a promising detector technology for detection of MeV 
gamma-rays~\cite{Takahashi01,Nakazawa04}.
Taking advantage of significant progress in CdTe technologies, we are 
developing a new generation of Compton telescopes, the SGD (Soft Gamma-ray 
Detector)~\cite{Takahashi02-NeXT,Takahashi03-SGD,Takahashi04-SGD,Takahashi04-NIM} onboard the NeXT (New X-ray Telescope) 
mission\cite{NeXT} proposed at ISAS (Institute of Space and Astronautical Science) as a successor of the Astro-E2.

The NeXT is optimized to study high-energy non-thermal processes: it will be a successor to the Astro-E2 mission, with much higher sensitivity in the energy range from 0.01 to 1 MeV.
The NeXT will have a dramatic impact on the sensitivity for source detection at hard X-ray and soft gamma-ray energies.
The broad bandpass will allow us to determine the range of energies of the radiating non-thermal particles, but the measurement beyond 511 keV provides the next crucial step towards the determination of the source structure: this is because the rate of $e^+/e^-$ pair production and their annihilation and thus the flux of the 511 keV line - depends critically on the compactness of the source. This is where the currently envisioned Soft Gamma-ray Detector (SGD) excels, with dramatically lower backgrounds as compared to previous instruments.

\section{Instrument Description}

\begin{figure}[bth]
   \centering
   \includegraphics[height=6cm]{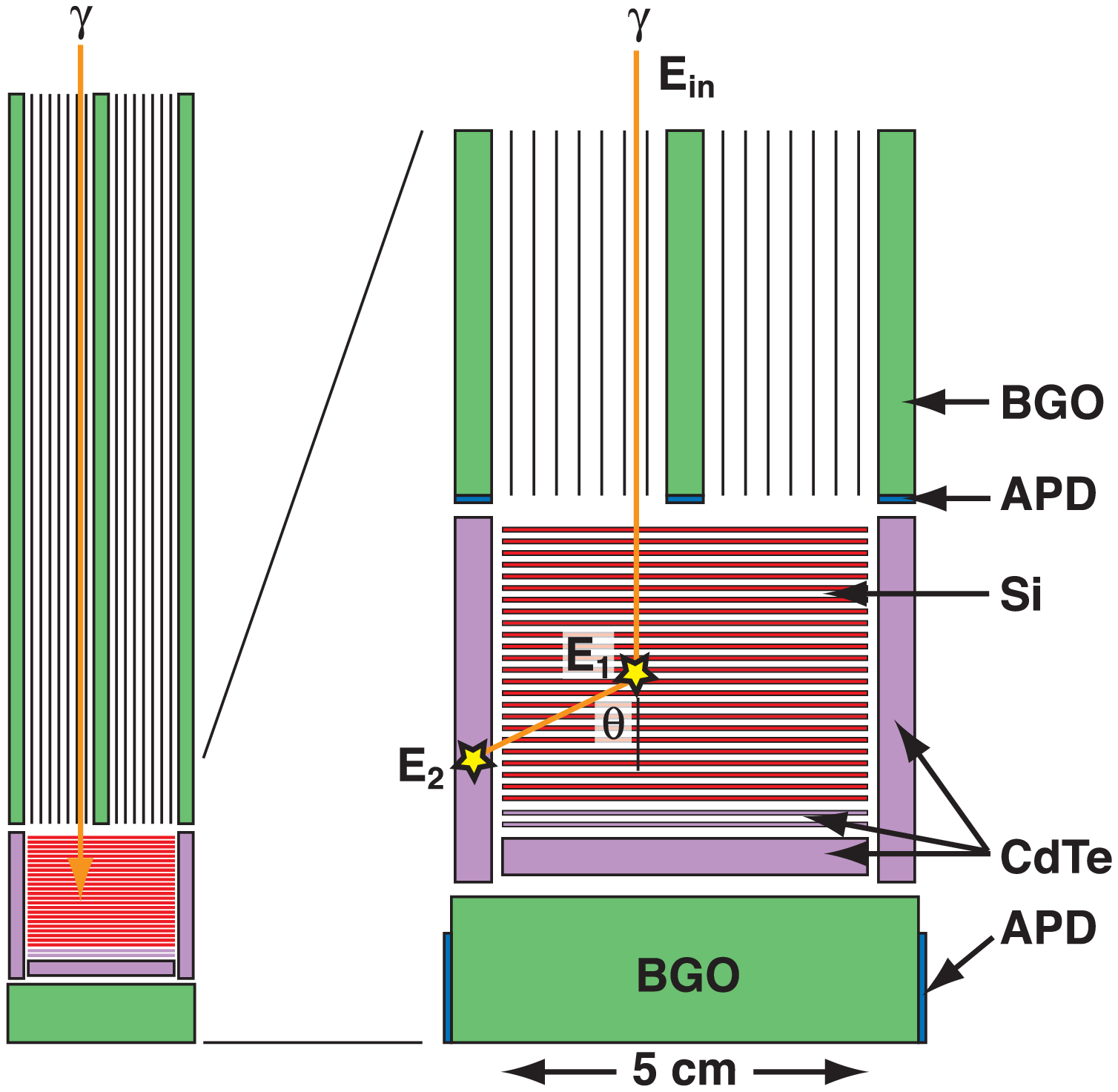}
   \caption[Conceptual drawing of a SGD detector unit.] 
   { \label{fig:SGD-unit} Conceptual drawing of a SGD detector unit.}
   \end{figure}

The NeXT/SGD is a Compton telescope with narrow FOV, which provides a constraint on Compton kinematics to enhance its background rejection capabilities.
The hybrid design of this module, illustrated in Fig.~\ref{fig:SGD-unit}, incorporates both silicon strips (to enhance response below $\sim$300 keV) and pixelated CdTe detectors.
The silicon layers also improve the angular resolution because of smaller effect from the finite momentum of the Compton-scattering electrons (Doppler broadening) than CdTe.
The Compton telescope consists of 24 layers of DSSDs (double-sided silicon strip detectors) and 2 layers of thin (0.5~mm) CdTe pixelated detectors surrounded by 5 mm thick CdTe pixelated detectors.
The DSSD strip pitch is chosen to be 0.4~mm to optimize the noise performance and power consumption.
The DSSD thickness is 0.5~mm mainly to avoid the high depletion voltage.
Thicker sensor is preferred from the view point of the higher interaction probability, however, the operation becomes increasingly difficult since the depletion voltage increases as the square of the thickness.
The pixel size of the CdTe detectors is $2\times 2$ \mmsq\ to optimize the angular resolution of the Compton kinematics and the number of channels required.
Thin CdTe detectors are placed underneath the DSSD layer to reduce the effect of the backgrounds as described in section~\ref{sect:BG}.
   \begin{figure}[bth]
   \centering
   \includegraphics[height=5cm]{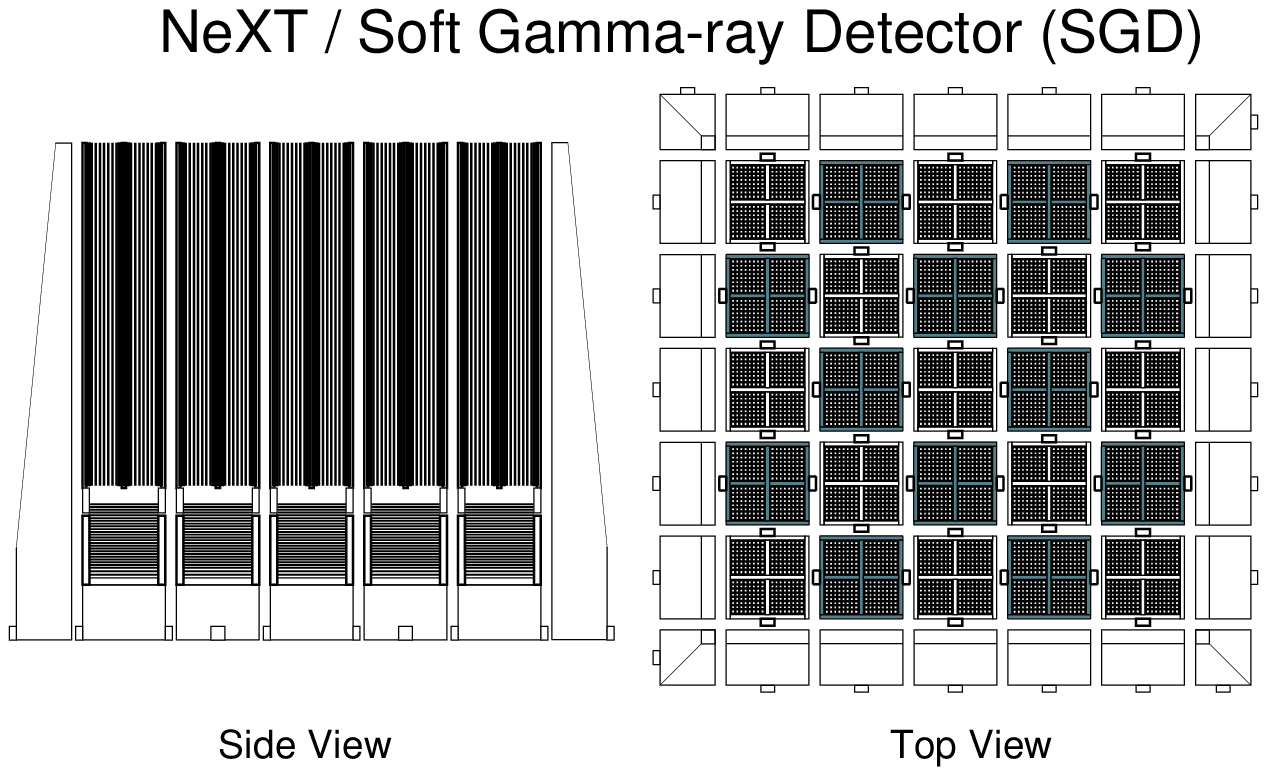}
   \caption[Conceptual drawing of SGD.] 
   { \label{fig:SGD-drawing} Conceptual drawing of the SGD instrument.}
   \end{figure} 

A copper collimator restricts the field of view of the telescope to 0.5\degree\ for low energy photons ($<$100~keV), which is essential to minimize the CXB (cosmic X-ray backgrounds).
A BGO (Bi${}_4$Ge${}_3$O${}_{12}$) active collimator defines the FOV to 4\degree\ for high energy photons.
A BGO bottom shield unit detects escaped photons and cosmic rays. 
Scintillation light from the BGO crystals is detected by avalanche photo-diodes (APDs) allowing a compact design and low energy threshold compared with phototubes.
The combination of a Compton telescope, BGO active collimator, a fine passive collimator constructed of copper and a BGO bottom shield forms a detector module.
These modules are then arrayed to provide the required area. Fig.~\ref{fig:SGD-drawing} shows a conceptual drawing of the SGD instrument, which consists of a $5\times 5$ array of identical detector modules surround by BGO shield units.

We require each SGD event to interact twice in the stacked detector, once by Compton scattering in the Si part, and then by photo-absorption in the CdTe part. 
Once the locations and energies of the two interactions are measured as shown in Fig.~\ref{fig:SGD-unit}, the Compton kinematics allows us to calculate the angle between the telescope axis and the incident direction of the event using the formula,
\begin{eqnarray}
\cos\theta &=& 1+\frac{m_ec^2}{E_2+E_1}-\frac{m_ec^2}{E_2},
\label{eq:kinematics}
\end{eqnarray}
where $\theta$ is the polar angle of the Compton scattering, and $E_1$ and $E_2$ are the energy deposited in each photon interaction.
The direction of the incident photon can be confined to be on the surface 
of a cone determined from $\theta$ and the two interaction positions.
The high energy resolution of the Si and CdTe devices help reduce the width of these ``Compton rings''.
We can determine the location of point sources as intersections of multiple rings. 
The angular resolution is limited to $\sim$8\degree\ at 100 keV due to the Doppler broadening (assuming that the energy resolution is better than 1.5 keV). 
In it important to reconstruct the order of the events correctly, and in order to do so, we can use the relation that the energy deposition by Compton scattering is always smaller than that of the photo absorption for energies below $E_{\gamma} = 256$ keV ($E_{\gamma} = m_e/2$).
This relation holds above this energy, if the scattering angle $\theta$ is smaller than $\cos^{-1}(1 - m_e / 2E_{\gamma}$). 
The major advantage of employing the Compton kinematics, however, is to reduce backgrounds. 
By having a narrow FOV, and by requiring the Compton ring of a valid aperture gamma-ray event to intersect with the FOV, we can reject most of background events. 
This dramatically reduces the background from radio-activation of the detector materials, which is a dominant background source in the case of the Astro-E2 HXD (Hard X-ray Detector)\cite{HXD,HXD-BG}.
Furthermore, we can eliminate Compton rings produced by bright sources located outside the FOV, which could produce significant background in some circumstances.
It is crucial to achieve low backgrounds since the photon sensitivity of the the HXD is limited by the backgrounds, not the effective area.
The Compton kinematics rely on the good angular resolution, which depends on the excellent energy resolution (better than 1.5 keV @ 60 keV) afforded by the detector material in the SGD module.

As a natural consequence of the Compton approach used to decrease backgrounds, the SGD module is quite sensitive to X/$\gamma$-ray polarization, thereby opening up a new window to study particle acceleration mechanism in astronomical objects.

\section{Background Suppression}\label{sect:BG}
Dominant background sources in the hard X-ray and soft gamma-ray bands are
\begin{itemize}
\item Source confusion and CXB (Cosmic X-ray Background).
\item Cosmic rays and their secondary particles produced by the interactions with the satellite.
\item Internal background due to the contamination of long-lived radio isotopes in the detector material\cite{HXD-BG}.
\item Internal background due to activation of the detector material resulting from the interaction with cosmic rays. SAA (South Atlantic Anomaly) proton dominates this effect in a low earth orbit planned for the NeXT mission.
\end{itemize}
   \begin{figure}[bth]
   \centering
   \includegraphics[height=5cm]{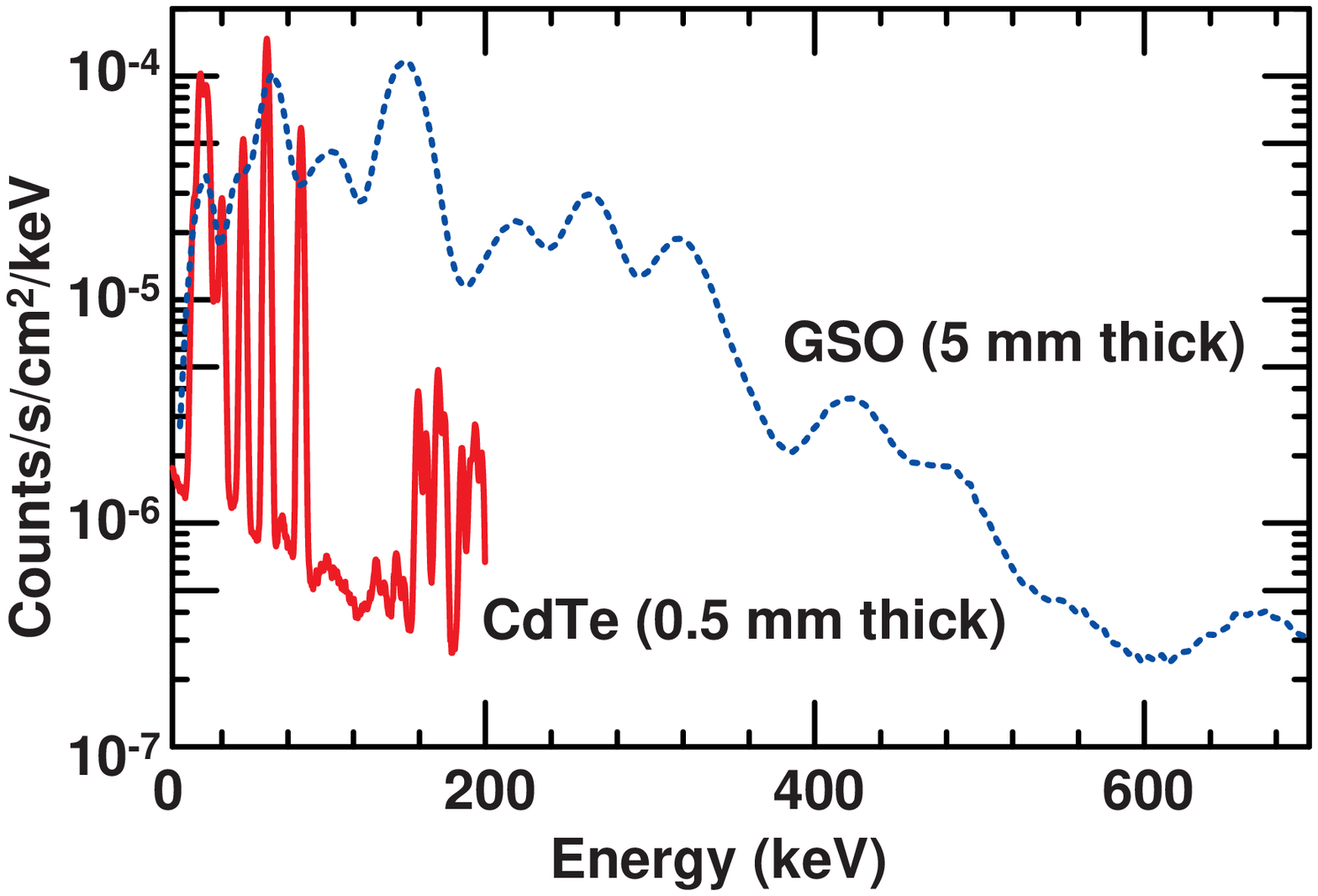}
   \caption[Spectra for 5~mm thick GSO and 0.5~mm thick CdTe activation backgrounds.] 
   { \label{fig:CdTe-activation} Spectra for GSO and CdTe internal backgrounds after activation.}
   \end{figure} 

The first source can be minimized by limiting the FOV.
The CXB is the dominant background source for the energy band less than 100~keV. To address this background term we implement a fine collimator, constructed of 50~\micron\ copper arranged to provide a 30 arc-minute field of view. 
The FOV in the higher energy region is 4\degree.

The backgrounds due to cosmic rays and secondary particles can be suppressed by the active BGO collimator and shield.
The effectiveness of this arrangement has been demonstrated by the Balloon-borne WELCOME detector\cite{Kamae92,Takahashi93}.

The radioisotope contamination will be a dominant background source for the Astro-E2 HXD, which is the predecessor of the SGD and utilizes a similar well-type active shield configuration. 
The major contaminant is from ${}^{152}$Gd in the GSO (Gd${}_2$SiO${}_5$) scintillator used in the HXD. 
The semiconductor devices we propose for use in the SGD are inherently free from any contamination. 
We still need to be concerned about the contamination in the BGO shields. 
Most of the background events emanating from the BGO shield are absorbed in the 5~mm thick CdTe detector and do not produce hits in the DSSD and thin CdTe layers.
The activation backgrounds from the CdTe can be minimized by limiting the volume of the CdTe detector illuminated by the incident photons.
Since the incident photons with energies less than 80~keV are absorbed in the
thin CdTe layers, the activation background is suppressed by a factor of $\sim$5 by requiring an interaction in either DSSD or thin CdTe layers.
Fig.~\ref{fig:CdTe-activation} shows the expected spectra of the internal backgrounds for 0.5~mm thick CdTe and 5~mm thick GSO after activation assuming the low earth orbit similar to the Astro-E2 sattellite.
These spectra are estimated from experimental results on the radioactivities induced by monoenergetic protons\cite{HXD-BG,Murakami03}
It clearly illustrates that the continuum background is reduced by more than an order of magnitude.

Further background suppression can be achieved by using the Compton events. 
In fact, most backgrounds with $E<100$~keV can be rejected by simply requiring two hits (albeit with an attendant reduction in the instrument effective area).
Applying the Compton kinematics and requiring consistency between the inferred incident photon direction and the FOV defined by the collimator will virtually eliminate the backgrounds from the bottom shield and CdTe detectors.
An albedo neutron background is also significantly suppressed by requiring two interactions and applying Compton kinematics and is negligible ($<10^{-8}$~\flux). 
The background from the BGO and copper collimator cannot be eliminated, but the source volume is small.

Incorporating background suppression by two-hit requirements and Compton kinematics into the internal background spectrum in Fig.~\ref{fig:CdTe-activation}, we estimate a background level of the SGD to be less than $5\times 10^{-7}$~\flux\ in the whole SGD energy band.
We plan to obtain more accurate background estimates using MGGPOD\cite{MGGPOD} simulation suite with a realistic mass model.

\section{Expected Performance}
The EGS4 Monte Carlo simulation package\cite{EGS4} with low energy extension\cite{EGS-KEK} is used to study the SGD performance.
A flat background level of $5\times 10^{-7}$~\flux\ is assumed throughout the studies unless otherwise indicated.
We performed several measurements and verified its predictions as described in section~\ref{sect:exp}.
In this section, EGS4 simulation studies of basic SGD performance are described.

\begin{figure}[bth]
   \centering
   \includegraphics[height=5cm]{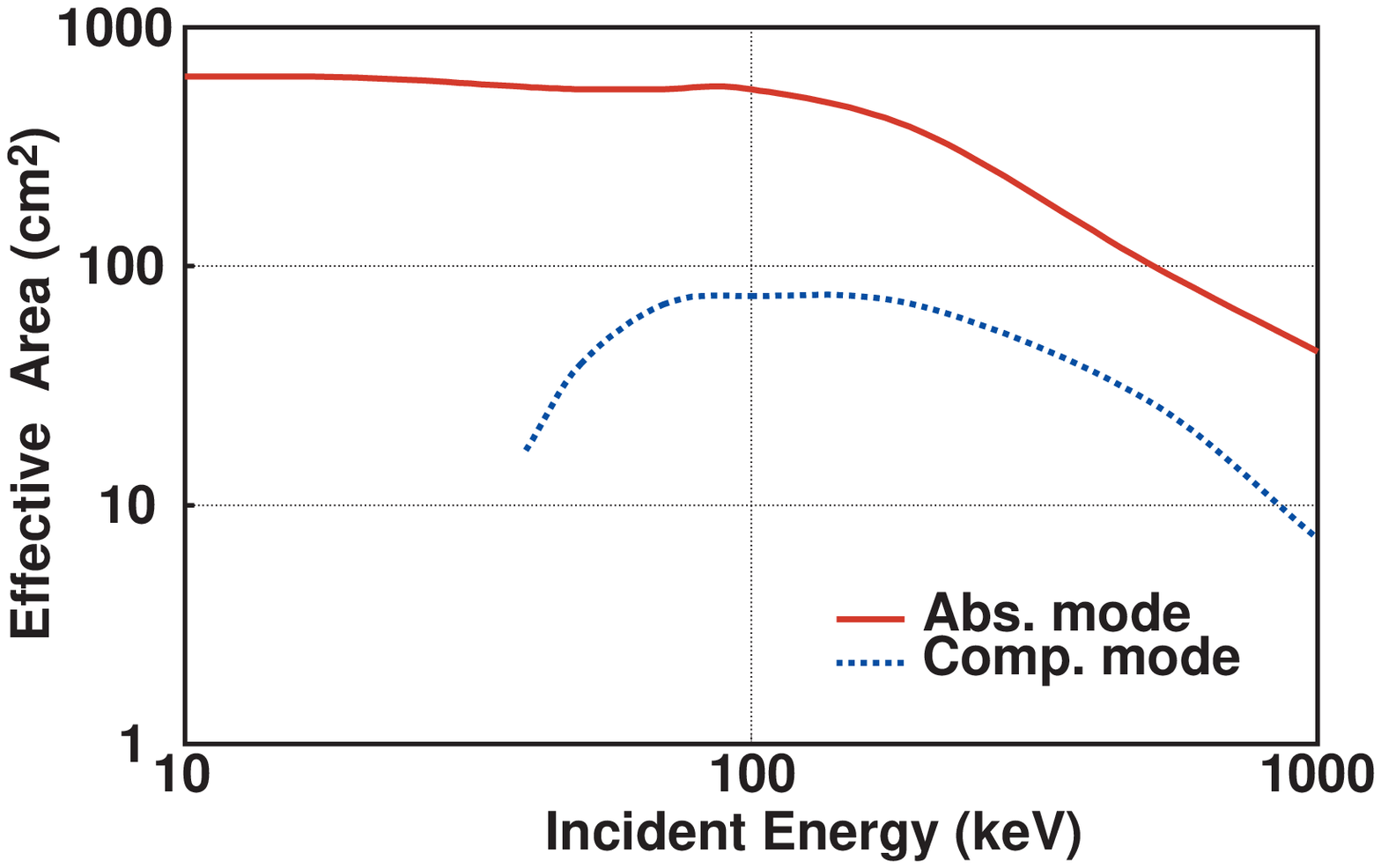}
   \caption[Effective area as a function of incident energy for the absorption mode (solid) and the Compton mode (dotted).] 
   { \label{fig:SGD-eff-area}Effective area as a function of incident energy for the absorption mode (solid) and the Compton mode (dotted).}
   \end{figure}

The effective area must be maximized in order to detect a sufficient number of photons for the observation of such faint objects enabled by the low background in a reasonable amount of time.
Fig.~\ref{fig:SGD-eff-area} shows the effective area as a function of the incident energy estimated using the EGS4 simulation. 
The solid curve represents the effective area with the absorption mode in which only total energy is measured without Compton kinematic constraint.
Incomplete events are still suppressed by BGO anti-shield detectors.
The dotted curve indicates the effective area with the Compton mode in which only events consistent with one Compton scattering and one photo absorption are accepted.
The difference between the absorption mode and the Compton mode is due to zero-Compton scattering events in lower energy region ($E<100$~keV) and multiple-Compton scattering events in higher energy region ($E>100$~keV).
Including multiple-Compton scattering events in the Compton mode can further improve the effective area in the higher energy region.

\begin{figure}[bth]
   \centering
   \includegraphics[height=5cm]{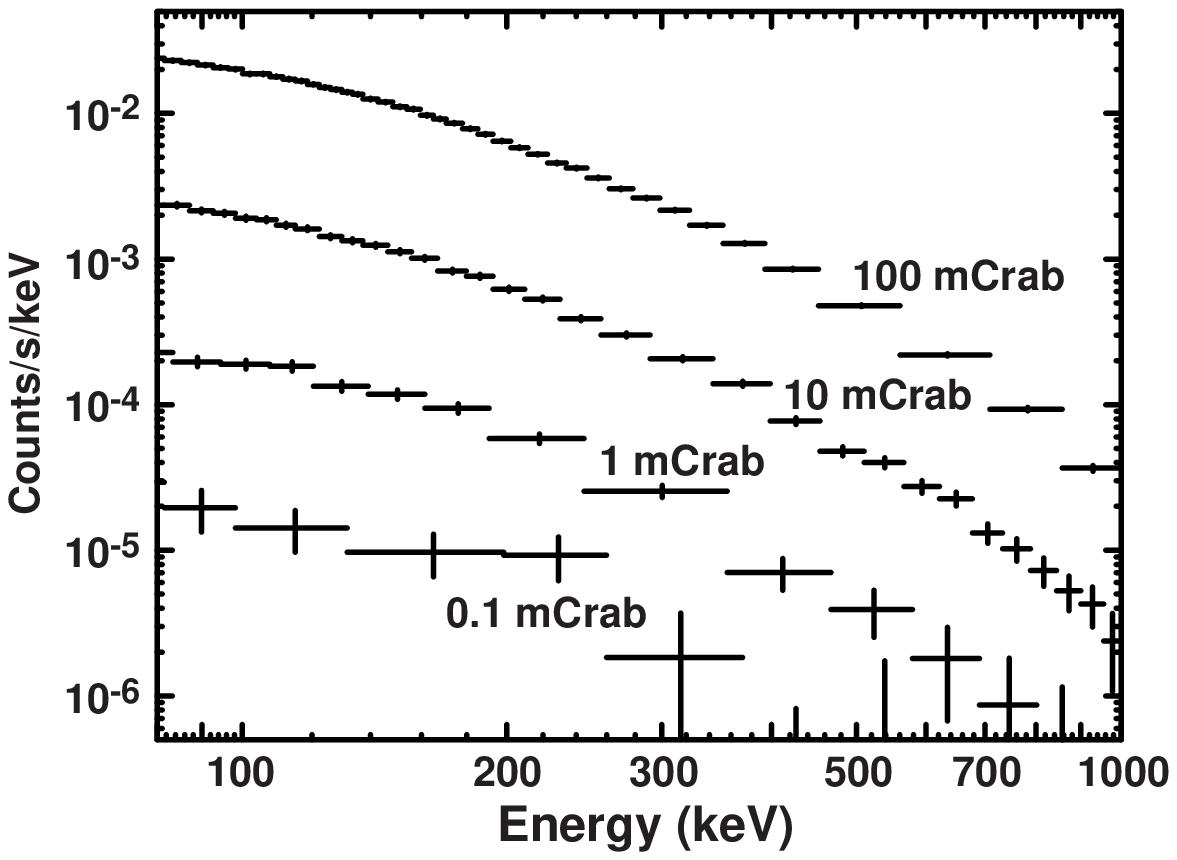}
   \caption[Expected energy spectrum for a 100~ks observation of 0.1--100~mCrab sources with the Compton mode.]
   { \label{fig:SGD-spectrum}Expected energy spectrum for a 100~ks observation of 0.1--100~mCrab sources with the Compton mode.}
   \end{figure}

\begin{figure}[bth]
   \centering
   \includegraphics[height=7cm]{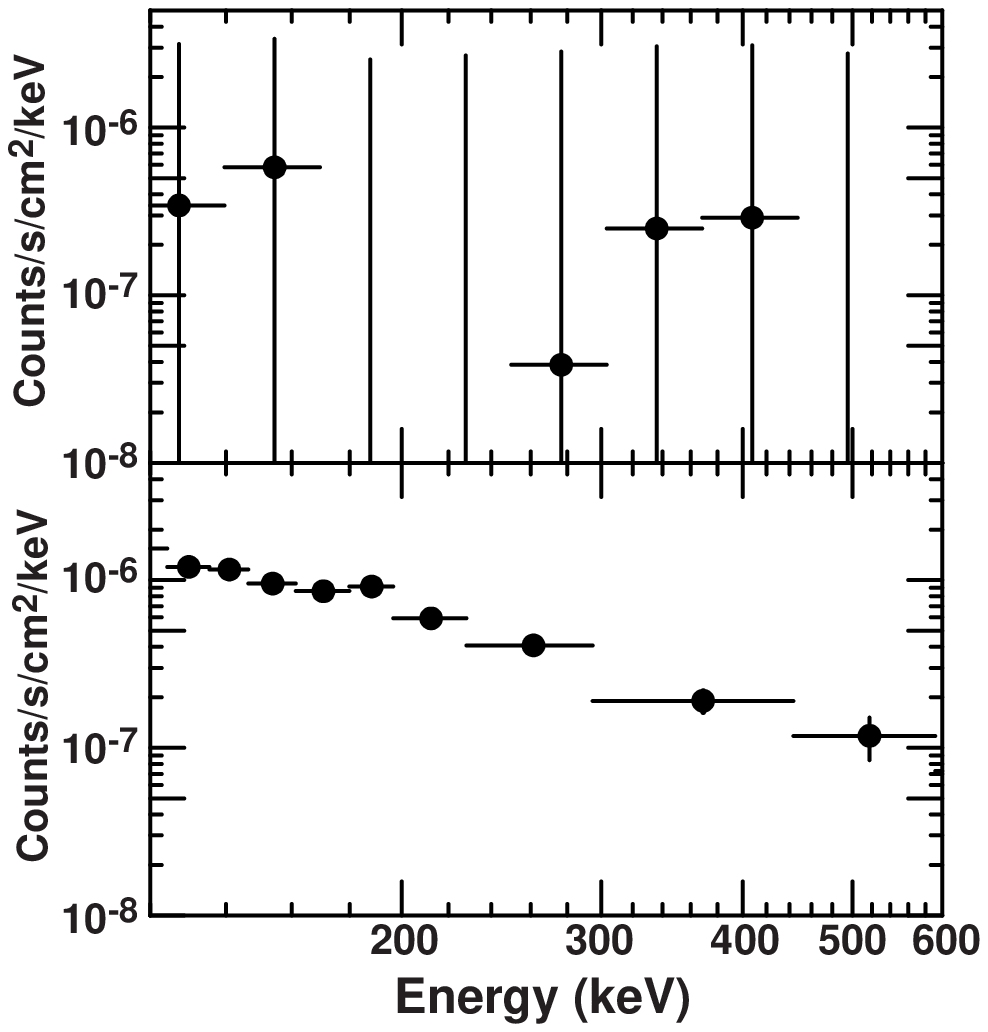}
   \caption[Comparison of the energy spectrum for a 100~ks observation of 1~mCrab source expected from the SGD Compton mode (bottom plot) and the instrument with the effective area of 3300~\cmsq\ and a background level of $5\times 10^{-4}$~\flux\ (top plot).]
   { \label{fig:SGD-spectrum-comp}Comparison of the energy spectrum for a 100~ks observation of 1~mCrab source expected from the SGD Compton mode (bottom plot) and the instrument with the effective area of 3300~\cmsq\ and a background level of $5\times 10^{-4}$~\flux\ (top plot).}
   \end{figure}

Fig.~\ref{fig:SGD-spectrum} shows the expected energy spectrum for a 100~ks observation of 0.1--100~mCrab sources with a spectral index of 1.7 using the Compton mode.
From errors on data points, we can conclude that the SGD can measure spectrum up to 800~keV for a 10~mCrab source and up to 300~keV for 1~mCrab source at better than 3$\sigma$ level.
Fig.~\ref{fig:SGD-spectrum-comp} compares the energy spectrum for a 100~ks observation of 1~mCrab source expected from the SGD Compton mode (bottom plot) and that for the instrument with the effective area of 3300~\cmsq\ (50 times the SGD effective area) and a background level of $5\times 10^{-4}$~\flux\ (top plot).
It clearly demonstrates the importance of the low background.

\begin{figure}[bth]
   \centering
   (a) 10~mCrab source\\
   \includegraphics[height=5cm]{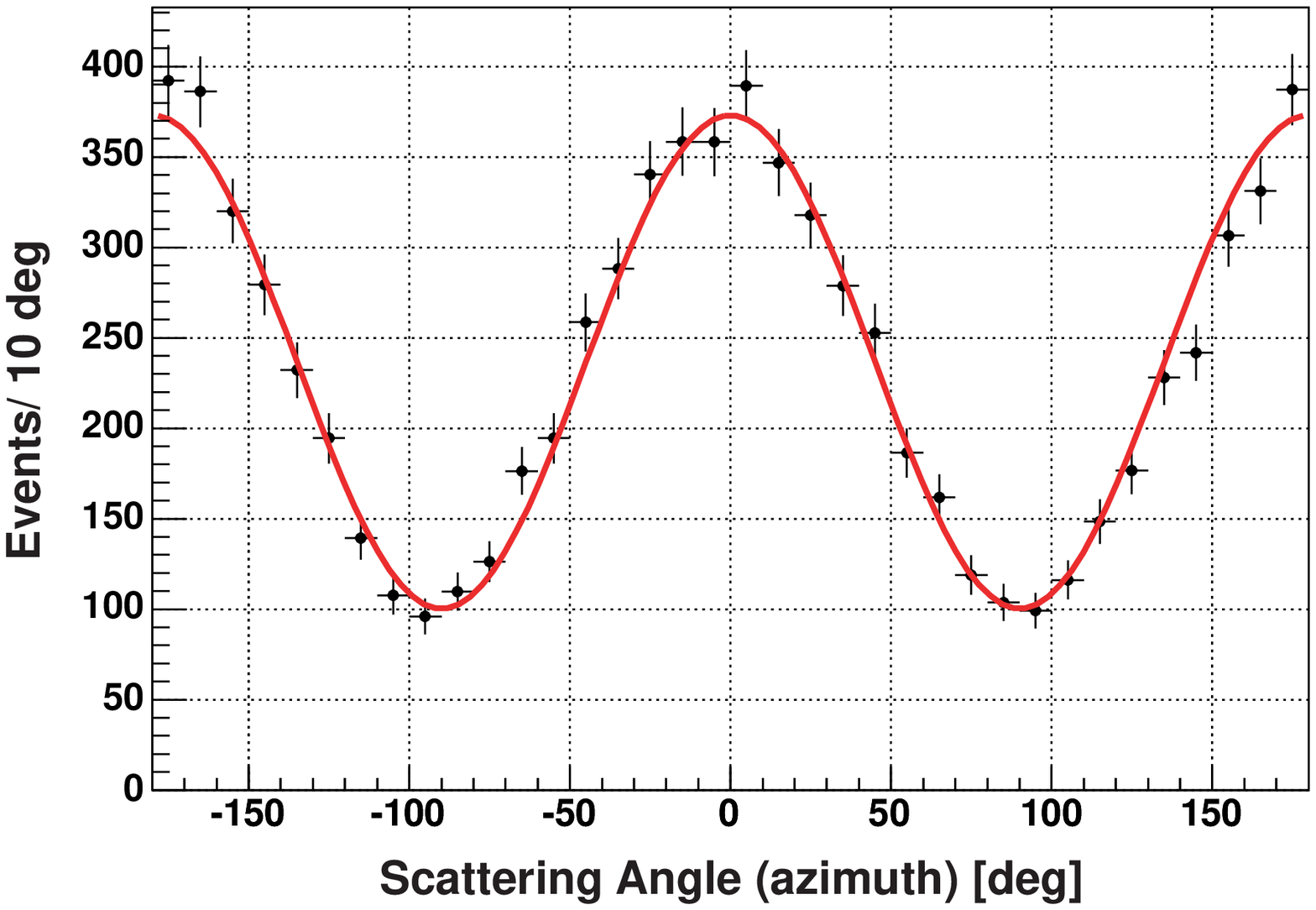}\\
   (b) 100~mCrab source \\
   \includegraphics[height=5cm]{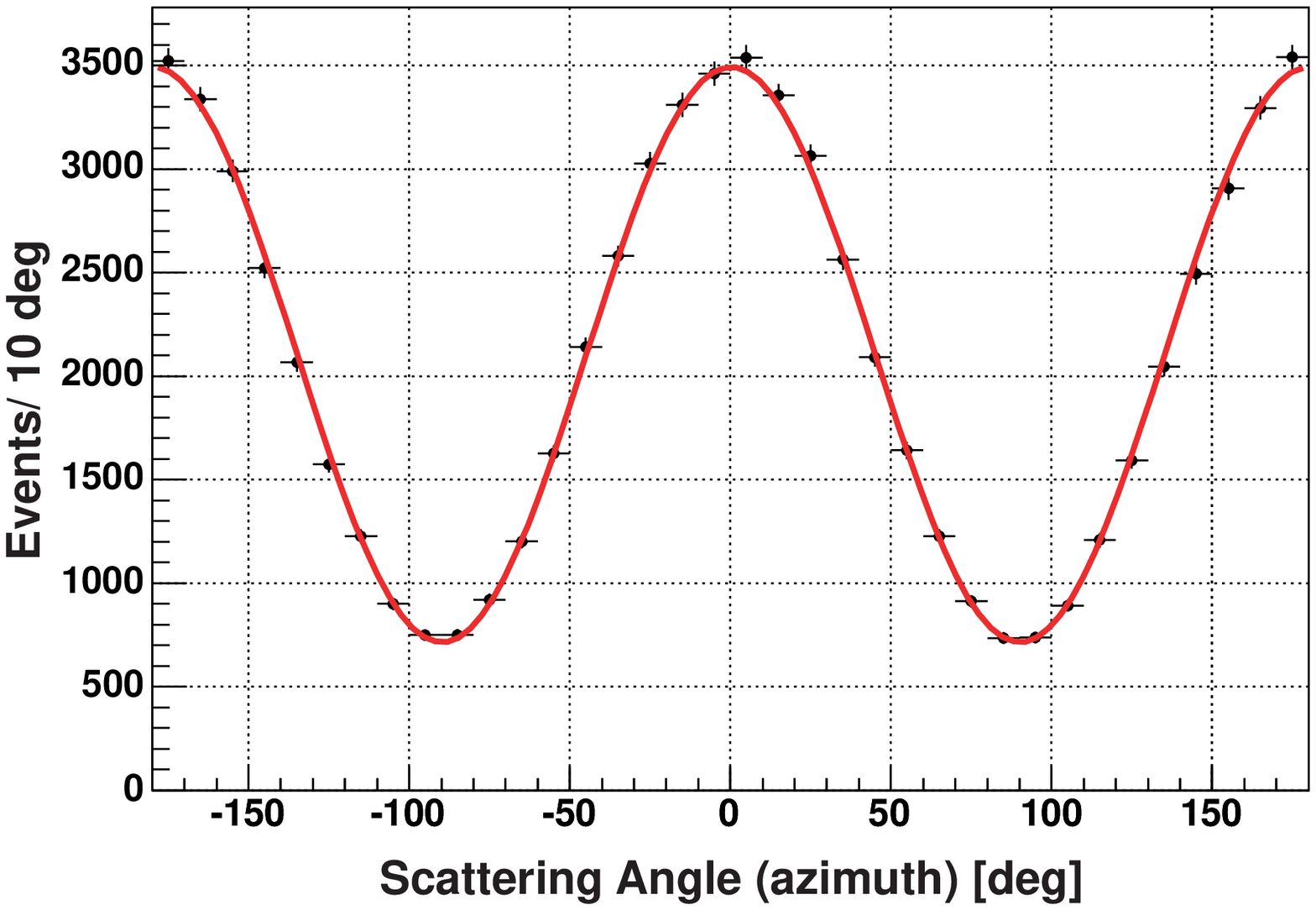} 
   \caption[Azimuth angle ($\phi$) distribution of the Compton scattering reconstructed for a 100~ks observation of 100 mCrab source with a 100\% polarization.]
   { \label{fig:pol-sensitivity} Azimuth angle ($\phi$) distribution of the Compton scattering reconstructed for a 100~ks observation of 10~mCrab and 100 mCrab sources with a 100\% polarization.}
   \end{figure}

Fig.~\ref{fig:pol-sensitivity} shows the azimuth angle ($\phi$) distribution of the Compton scattering reconstructed for a 100~ks observation of 10~mCrab and 100~mCrab sources with a 100\% linear polarization.
Events with observed energy above 300~keV are not used to suppress backgrounds.
The distribution is fit to a formula, $AVG\cdot (1+Q\sin2(\phi-\chi))$, where $AVG$ is the average flux per bin, $Q$ is the modulation factor which should be proportional to the polarization, and $\chi$ is the polarization phase and perpendicular to the polarization vector.
The fit yields the modulation factor of $57.6\pm1.3$\% for a 10~mCrab source and $66.0\pm0.4$\% for a 100~mCrab source.
These results correspond to the 5$\sigma$ polarization sensitivity of 11.3\% and 3.2\%, respectively.
Detailed studies of the SGD polarization performance can be found in \cite{Tajima04-POL}.

\section{Detector Performance Verification}\label{sect:exp}

\subsection{Energy Resolution}
\begin{figure}[tbh]
   \centering
   \includegraphics[height=4.6cm]{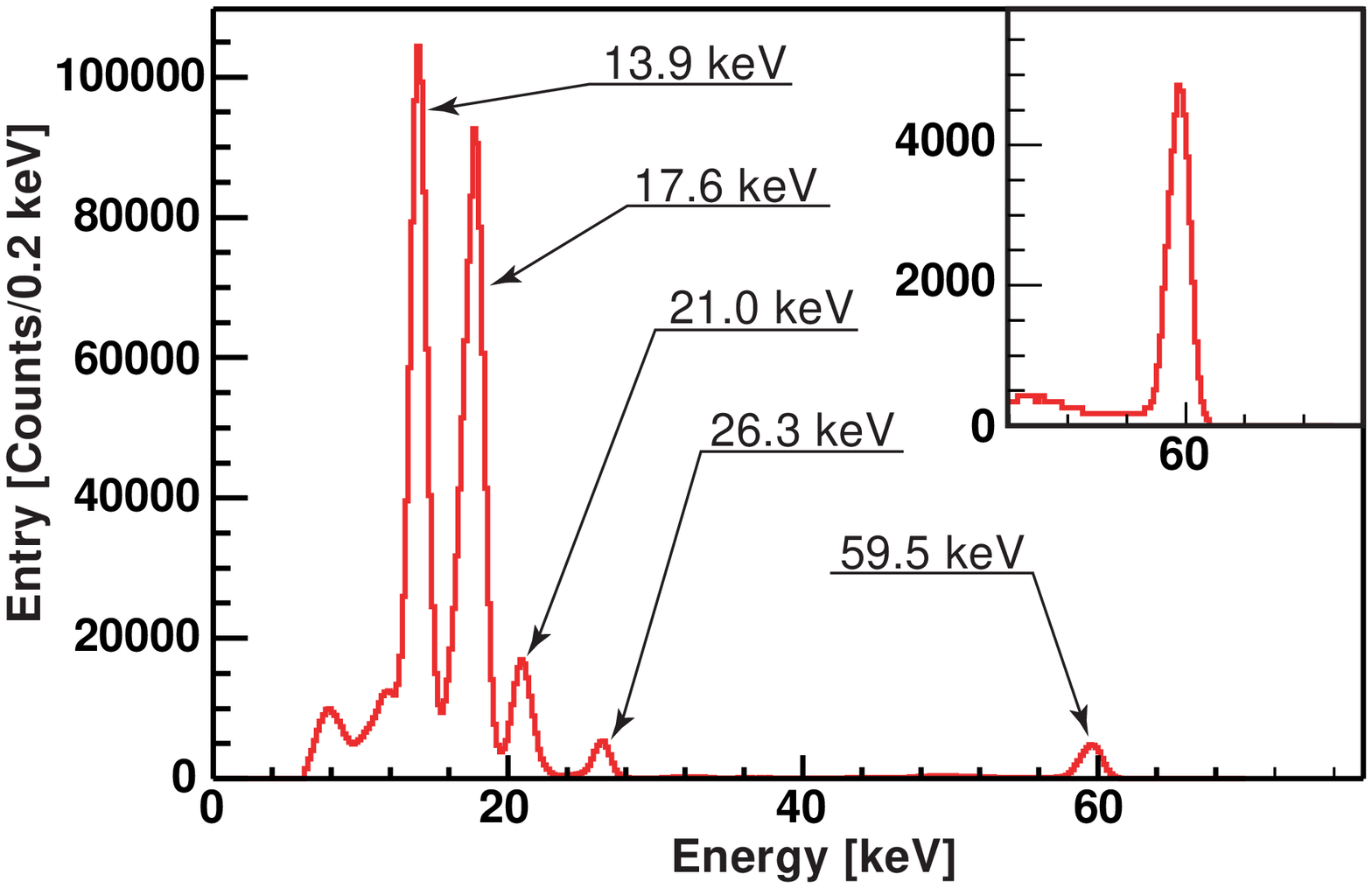}
   \caption[] 
   { \label{fig:Am-SSD} Energy spectrum of ${}^{241}$Am for a DSSD measured at 0\degC. The 60~keV peak is magnified in the vertical axis in inset plot.}
   \end{figure}

\begin{figure}[tbh]
   \centering
   \includegraphics[height=4.6cm]{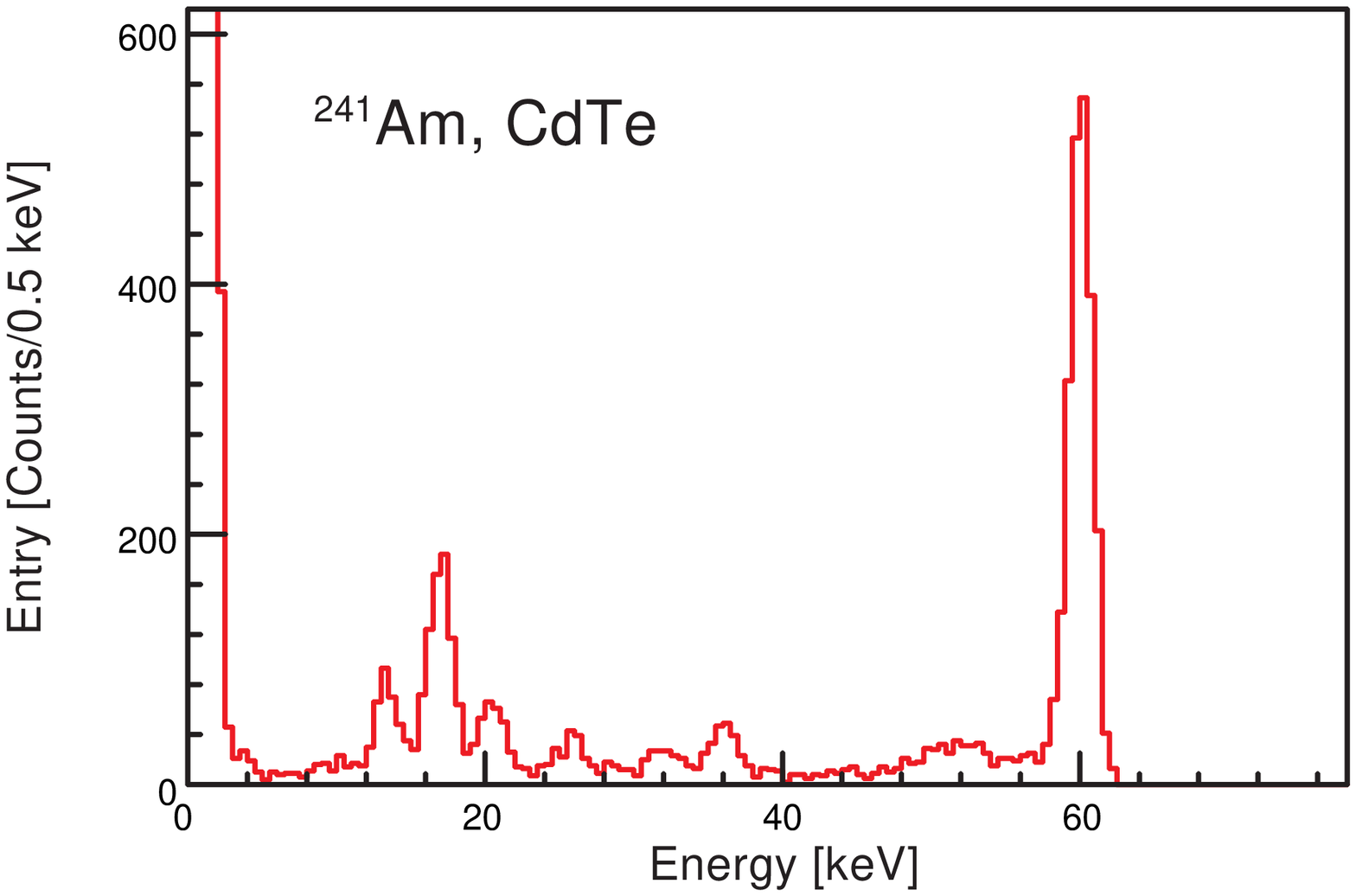} 
   \caption
   { \label{fig:CdTe-spectrum} The ${}^{241}$Am energy spectrum for 
a CdTe pixel detector. }
   \end{figure}

A fine energy resolution is one of the key requirements for the SGD to retain a low background level with Compton kinematics.
The energy resolution around 1~keV is required to approach the physical limit of the angular resolution due to Doppler broadening.
We have developed a low noise front-end ASIC (Application-Specific Integrated Circuit), VA32TA, to realize this goal.
We have developed prototype modules for a low noise DSSD system in order to evaluate noise sources.
To keep the strip yield close to 100\% and eliminate polysilicon bias 
resistor (a possible noise source), the DSSD does not employ an integrated 
AC capacitor.
We have produced 0.3~mm thick DSSDs with a strip length of 2.56~cm, 
a strip pitch of 0.4~mm and a strip gap of 0.1~mm.
The C-V measurement indicates a depletion voltage of 65~V, 
therefore the following measurements are performed with a 70~V bias voltage.  
The leakage current is 0.5~nA/strip at 20\degC\ and 0.05~nA/strip at 0\degC.
The strip capacitance is measured to be $6.3\pm0.2$~pF.
Intrinsic noise performance with a silicon strip detector is measured 
to be 1.0~keV (FWHM) at 0\degC, which is in good agreement with the analytically calculated noise value of 0.9~keV\cite{Tajima02,Tajima04}.
The energy resolution for $\gamma$-rays is investigated using the 14.4~keV and 122.06~keV $\gamma$-ray lines from ${}^{57}$Co and the 59.54~keV $\gamma$-ray line from ${}^{241}$Am.
   \begin{figure}[tbh]
   \centering
   \includegraphics[height=4.6cm]{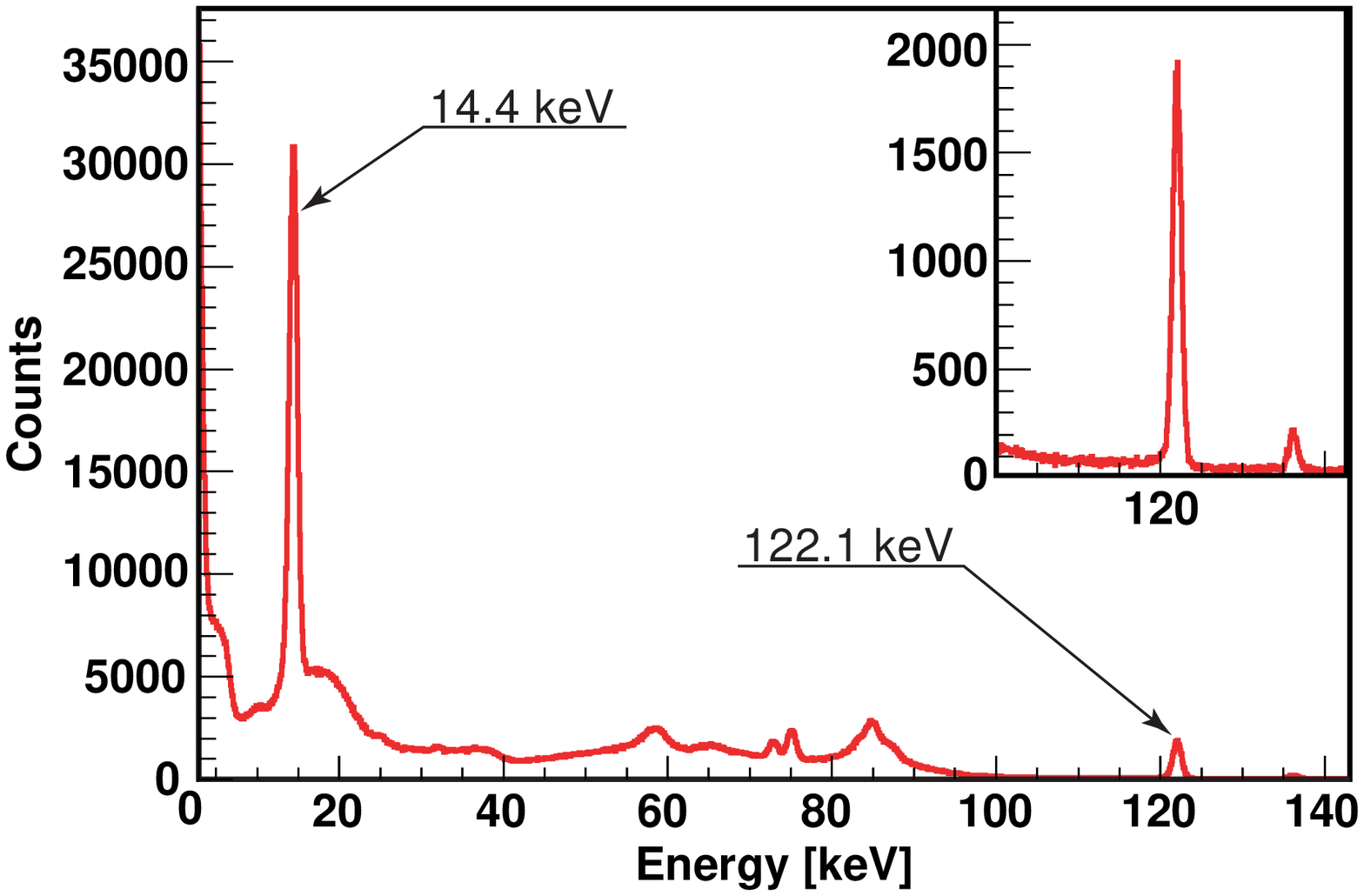}
   \caption[] 
   { \label{fig:Co-SSD} Energy spectrum of ${}^{57}$Co for a DSSD measured at $-10$\degC. The 122~keV peak is magnified in the vertical axis in inset plot.}
   \end{figure} 
Fig.~\ref{fig:Co-SSD} shows the sum of the energy spectrum from ${}^{57}$Co for all channels at $-10$\degC.
Fits yield energy resolutions of 1.1~keV (FWHM) for the 14~keV line and 1.3~keV for the 122~keV line\cite{Fukazawa04}.
Fig.~\ref{fig:Am-SSD} shows the energy spectrum from ${}^{241}$Am at 0\degC.
A fit yields an energy resolution of 1.3 keV\cite{Tajima02,Tajima04,Fukazawa04}.
We expect the energy resolution with the full size DSSD to be about 10\% worse than this result due to large strip capacitance load.
The VA32TA performance for a CdTe detector is also evaluated using an $8\times
8$ array of $2\times2$~\mmsq\ CdTe pixels.
The thickness of the detector is 0.5~mm.
Each pixel is connected to VA32TA via a fan-out board.
Capacitance and leakage current of each pixel are 1~pF and a few pA at 0\degC.
Low leakage current is realized by employing a guard ring to absorb leakage 
current from the detector edge~\cite{Nakazawa04}.
Noise is measured to be 1.5 keV (FWHM).
Fig.~\ref{fig:CdTe-spectrum} shows the ${}^{241}$Am energy spectrum 
at a bias voltage of 600~V.
We obtain an energy resolution of 1.7 keV (FWHM)~\cite{Mitani04,Tajima04}.

\begin{figure}[bth]
   \centering
   \includegraphics[height=5cm]{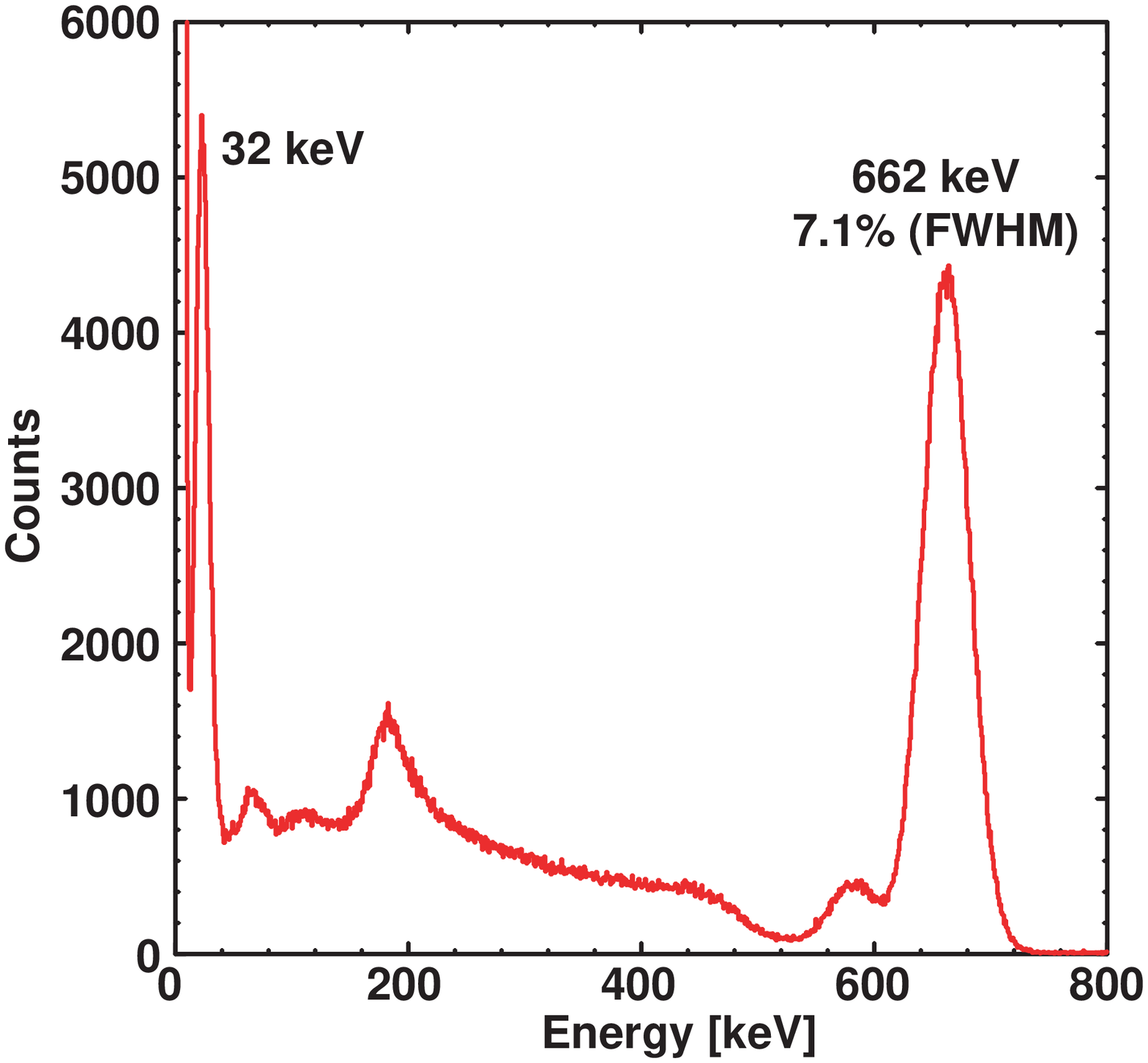}
   \caption[] 
   { \label{fig:BGO-20} The ${}^{137}$Cs energy spectrum for a $10\times 10\times 10$~mm${}^3$ BGO crystal with an APD readout at $-20$\degC.}
   \end{figure}

\begin{figure}[bth]
   \centering
   \includegraphics[height=5cm]{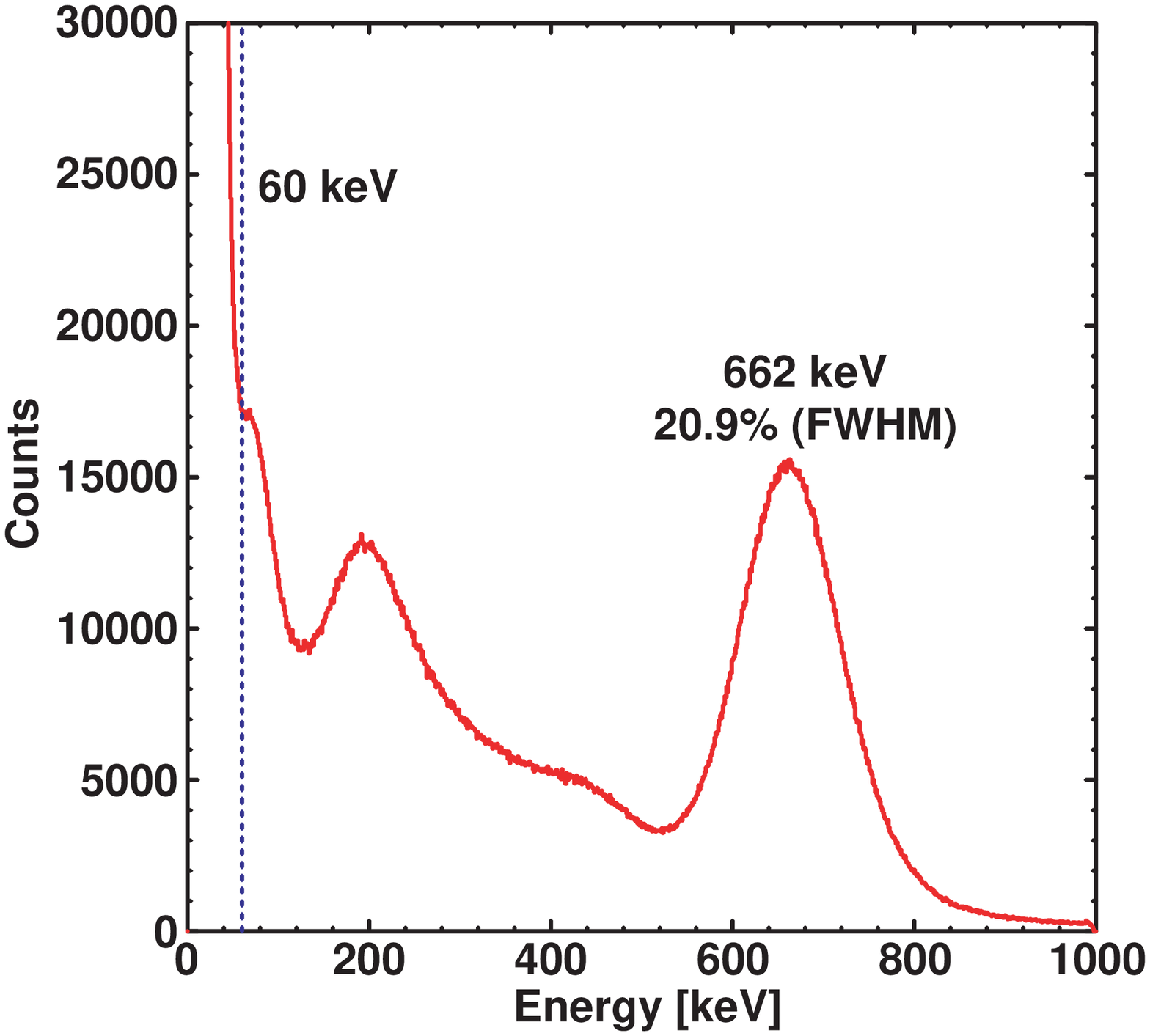}
   \caption[] 
   { \label{fig:largeBGO} The ${}^{137}$Cs energy spectrum for a $300\times 48\times 3$~mm${}^3$ BGO crystal with an APD readout at $-15$\degC.}
   \end{figure}

The APD readout of BGO is a crucial part of the SGD detector development since it is extremely difficult to readout BGO collimators with phototubes due to mechanical constraints.
Higher quantum efficiency of the APD will also contribute to lower the energy threshold of BGO\cite{Ikagawa03,Nakamoto04}.
A large area reverse-type APD ($10\times10$~\mmsq), S8664-1010N, has been developed by Hamamatsu Photonics.
The reverse-type APD can obtain sufficient internal gain at relatively lower bias voltage and low leak current.
The S8664-1010N achieves the internal gain of 50 with the bias voltage of $\sim$380~V at $-20$\degC\ while keeping the leak current at $\sim$60~pA\cite{Ikagawa04}.
We have evaluated the performance of this APD attached to BGO crystals.
Fig.~\ref{fig:BGO-20} and Fig.~\ref{fig:largeBGO} show the ${}^{137}$Cs energy spectrum with a $10\times 10\times 10$~mm${}^3$ BGO block at $-20$\degC\ and a $300\times 48\times 3$~mm${}^3$ BGO plate at $-15$\degC, respectively.
The small BGO is attached directly to the APD while the large BGO is attached to the APD via a light guide.
Energy resolutions are 7.1\% (FWHM) for the small BGO and 20.9\% for the large BGO\cite{Ikagawa04}.
Minimum detectable energies are 11~keV for the small BGO and 60~keV for the large BGO.
Preliminary measurements with an array of nine 3-mm diameter APDs achieved minimum detectable energies of $\sim$40~keV at $-20$\degC\ and $\sim$20~keV at $-40$\degC.
We plan to develop an APD which closely matches the cross section of the large BGO plate to further improve the performance, namely to enable the operation of the instrument above $-10$\degC.

\subsection{Compton Reconstruction}
\begin{figure}[bth]
   \centering
   \includegraphics[height=5cm]{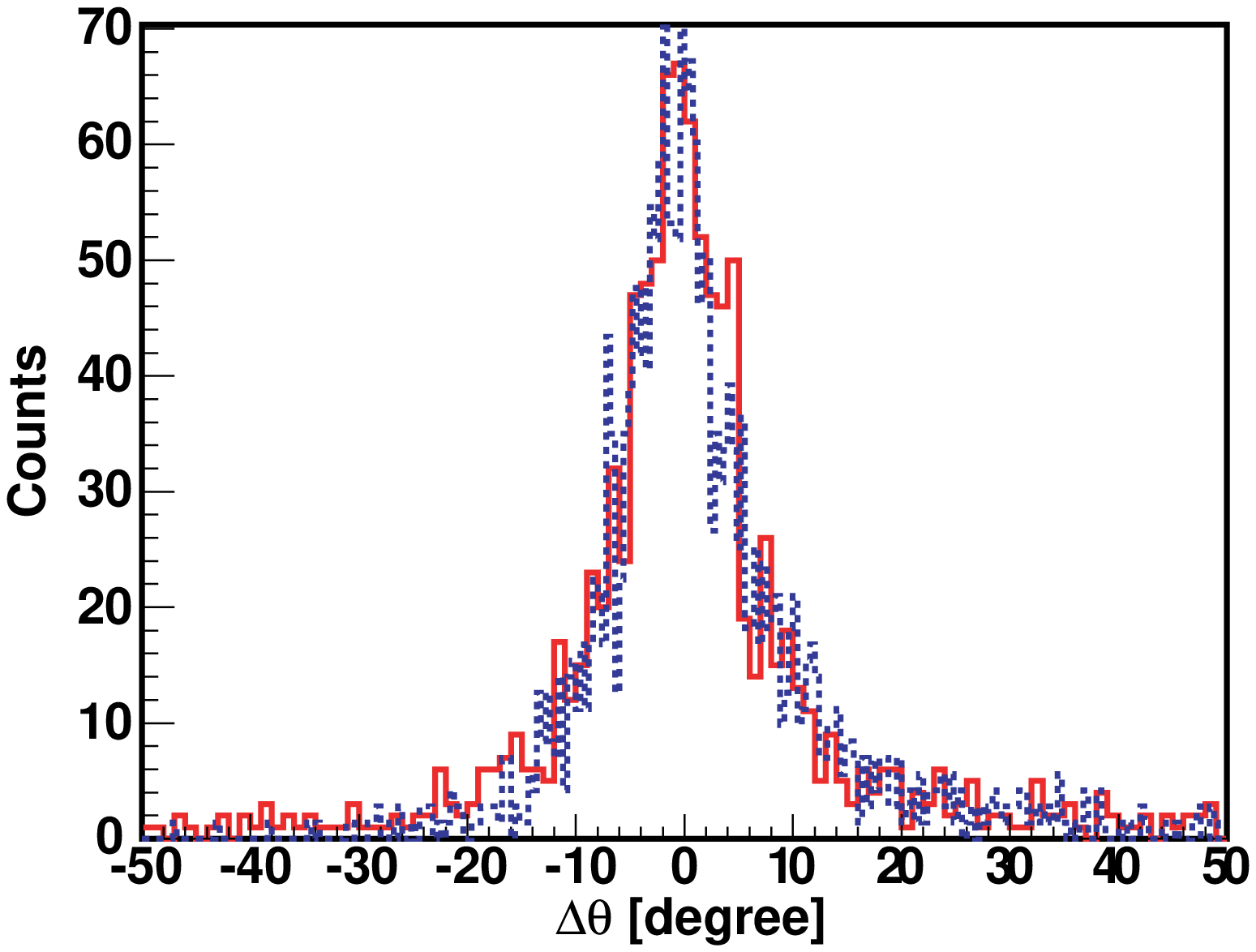}
   \caption[] 
   { \label{fig:Comp-angle} Distribution of the difference between the Compton scattering angle calculated from the energy deposits and the geometrical angle  obtained using a 122~keV $\gamma$-ray line.}
   \end{figure}

We have assembled a small prototype Compton telescope that consists of two DSSD systems to evaluate the angular resolution of the Compton reconstruction\cite{Fukazawa04}.
Two DSSDs are placed in parallel with a separation of 6.7~mm and exposed to a ${}^{57}$Co source at a distance of 56~mm with a 2~mm diameter Pb collimator.
Events that are consistent with two interactions, which are predominantly events with one Compton scattering and one photo-absorption, are selected by requiring that the sum of the energies of two hits in two DSSDs is within 2$\sigma\approx 1.3$~keV.
In these events, the Compton scattering angle can be calculated both from two energy deposits using Compton kinematics and from two hit positions and the known source position.
Fig.~\ref{fig:Comp-angle} shows $\Delta\theta$, the difference between the Compton scattering angle calculated from the energy deposits and the geometrical angle obtained using a 122~keV $\gamma$-ray line.
The solid histogram represents the experimental result while the dotted histogram shows the result from Geant4 Monte Carlo (MC) simulation\cite{G4} with G4LECS for Compton scattering process\cite{Kippen}.
They are in a good agreement and yield an angular resolution of 8\degree\cite{Fukazawa04}.
The MC simulation indicates that Doppler broadening is the dominant component (7\degree) and the energy resolution has already reached adequate level.
   \begin{figure}[bth]
   \centering
   \includegraphics[height=5cm]{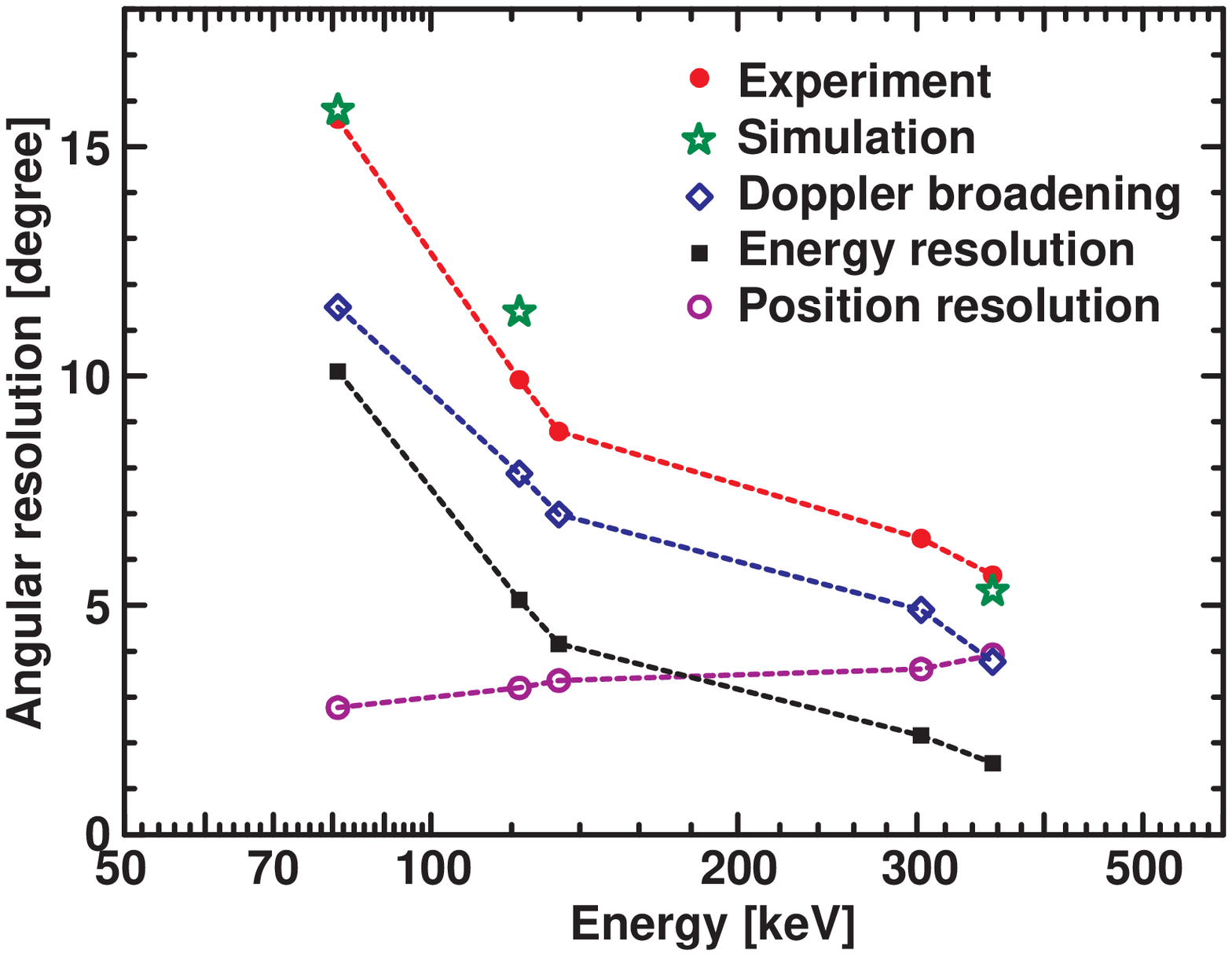}
   \caption[] 
   { \label{fig:ang_res-energy} Angular resolution as a function of energy for the experimental data and the MC simulation. Contributions from Doppler broadening, energy resolution and position resolution are also displayed. }
   \end{figure} 
   \begin{figure}[bth]
   \centering
   \includegraphics[height=5cm]{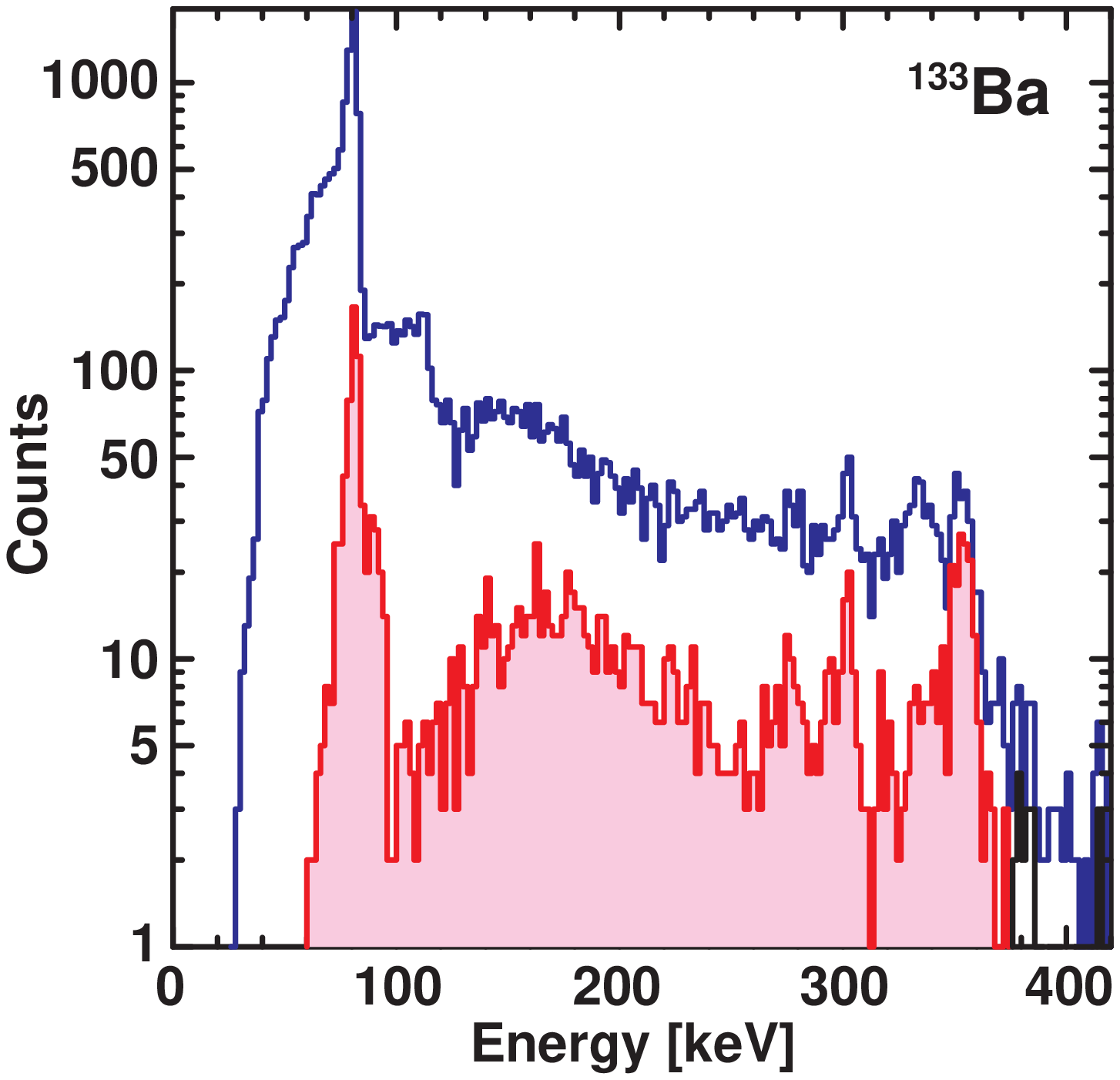}
   \caption[] 
   { \label{fig:BG-rejection} ${}^{133}$Ba spectra with (hatched histogram) and without the background rejection by Compton kinematics.}
   \end{figure} 
   \begin{figure}[bth]
   \centering
   (a) Experimental result\\
   \includegraphics[height=5cm]{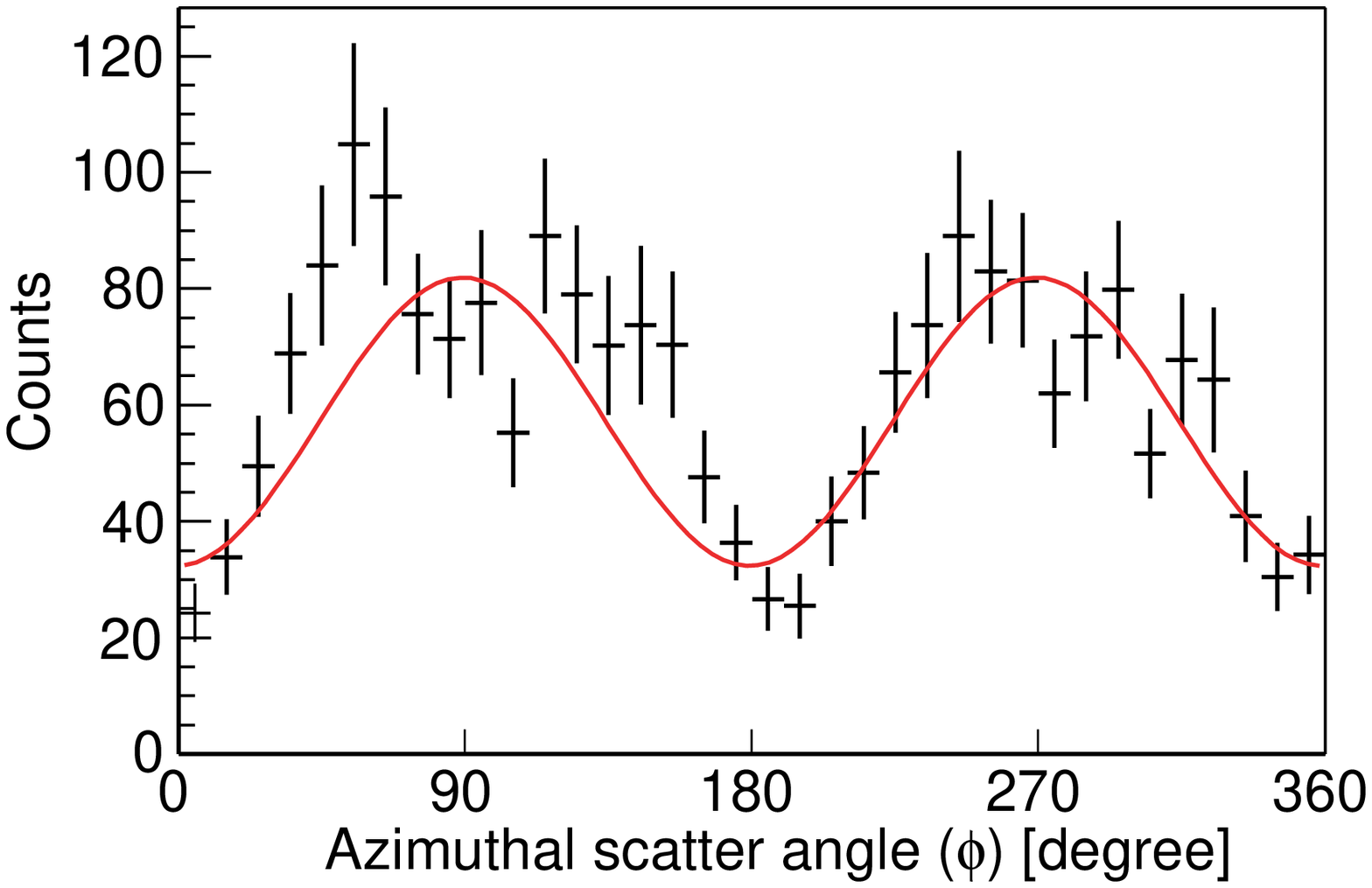}\\   
   (b) EGS4 simulation \\
	\includegraphics[height=5cm]{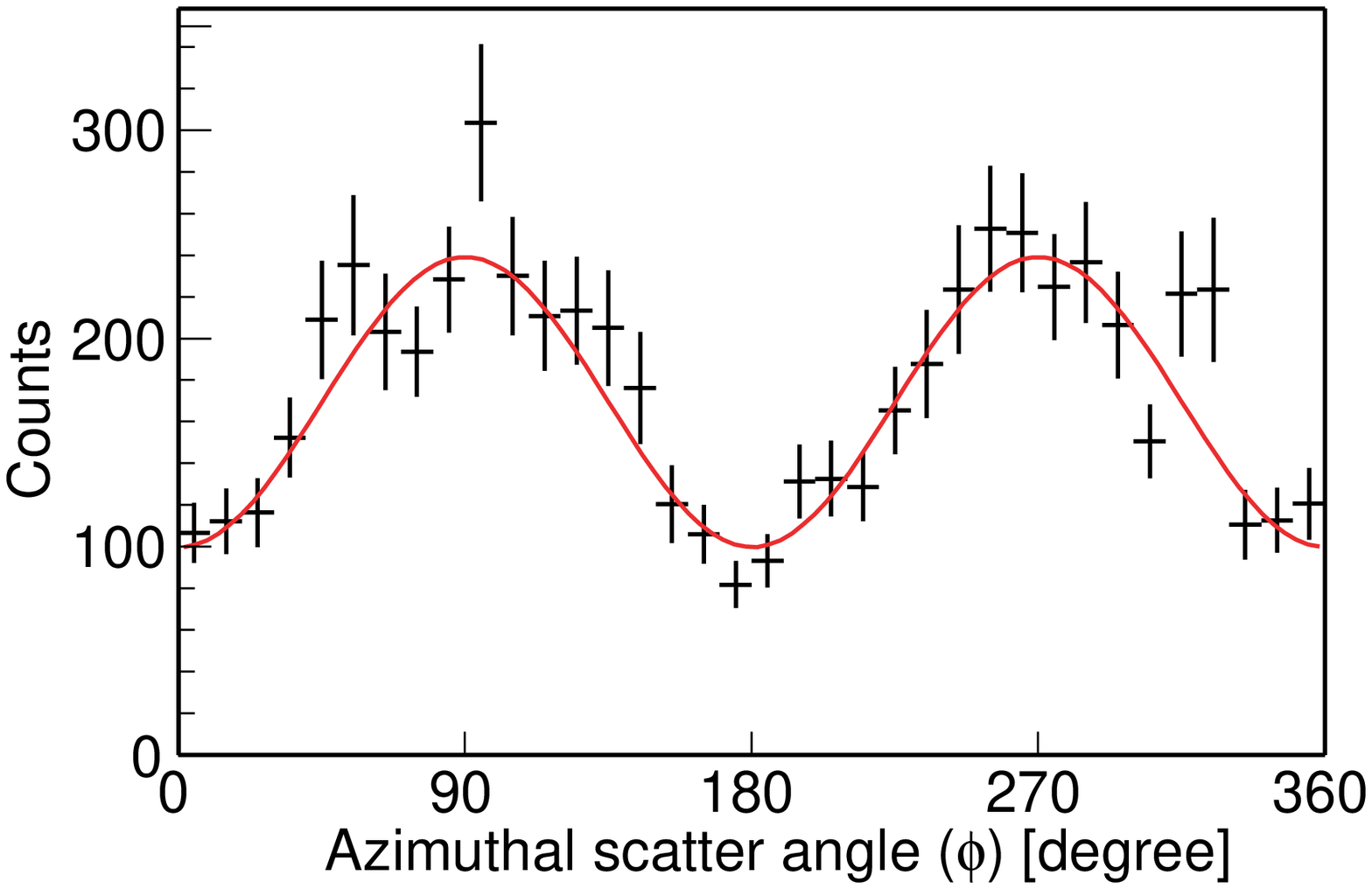}
   \caption[(a) Azimuth angle distribution of the Compton scattering obtained using a Compton telescope consists of a DSSD and CdTe pixel detectors with 177~keV polarized photon beam at SPring-8. (b) Corresponding distribution by EGS4 simulation.] 
   { \label{fig:sp8-phi} (a) Azimuth angle distribution of the Compton scattering obtained using a Compton camera consisting of a DSSD and CdTe pixel detectors with 177~keV polarized photon beam at SPring-8. (b) Corresponding distribution by EGS4 simulation.}
   \end{figure} 

Further studies are performed using a prototype Compton telescope that consists of a DSSD and two CdTe detectors\cite{Tanaka04}.
The DSSD and the CdTe detectors are placed in parallel with a distance of 12.5~mm.
In this measurement, the telescope is exposed to ${}^{57}$Co (122 and 136~keV) and ${}^{133}$Ba (81~keV, 303~keV and 356~keV) sources to study the performance in wide energy range.
Events with one Compton scattering in the DSSD and one photo-absorption in the CdTe are selected using Compton kinematics in a similar manner as above.
Fig.~\ref{fig:ang_res-energy} shows the angular resolution as a function of the energy\cite{Tanaka04}.
It also shows the Geant4 MC simulation estimates for the total angular resolution and contributions from Doppler broadening, energy resolution and position resolution.
The Doppler broadening effect is the dominant source of the degradation of the angular resolution up to 300~keV.
The position resolution becomes significant in higher energy.
Using the same data set, the effect of the background rejection by Compton kinematics is also studied.
First, the energy spectrum is obtained for events with one hit in the DSSD and one hit in the CdTe.
Then, the background rejection is applied by requiring $|\Delta\theta|<16$\degree.
Fig.~\ref{fig:BG-rejection} shows the ${}^{133}$Ba spectra with (hatched histogram) and without the background rejection by Compton kinematics\cite{Tanaka04}.
It clearly demonstrates the positive effect of this technique.
Note that the background rejection criteria are not tuned and can be improved significantly, in particular in the low energy region.
Detailed description of Compton reconstruction studies can be found elsewhere\cite{Tajima03,Fukazawa04,Tanaka04,Watanabe04}.

In order to evaluate the polarization performance of the prototype Compton telescope, we have carried out a beam test at SPring-8 photon factory in Japan\cite{Mitani04}.
Fig.~\ref{fig:sp8-phi} (a) shows the azimuth angle distribution of the Compton scattering obtained using 177~keV photon beam with a 100\% linear polarization.
Fig.~\ref{fig:sp8-phi} (b) shows the corresponding distribution simulated by the EGS4 MC simulation.
A fit on the experimental measurement yields the modulation factor of $43\pm3$\% which is in a good agreement with the EGS4 result of $41\pm2$\%\cite{Mitani04}.
These results validate our implementation of the EGS4 simulation at 10\% level.

\section{Conclusions}
\begin{figure}[bth]
   \centering
   \includegraphics[height=7cm]{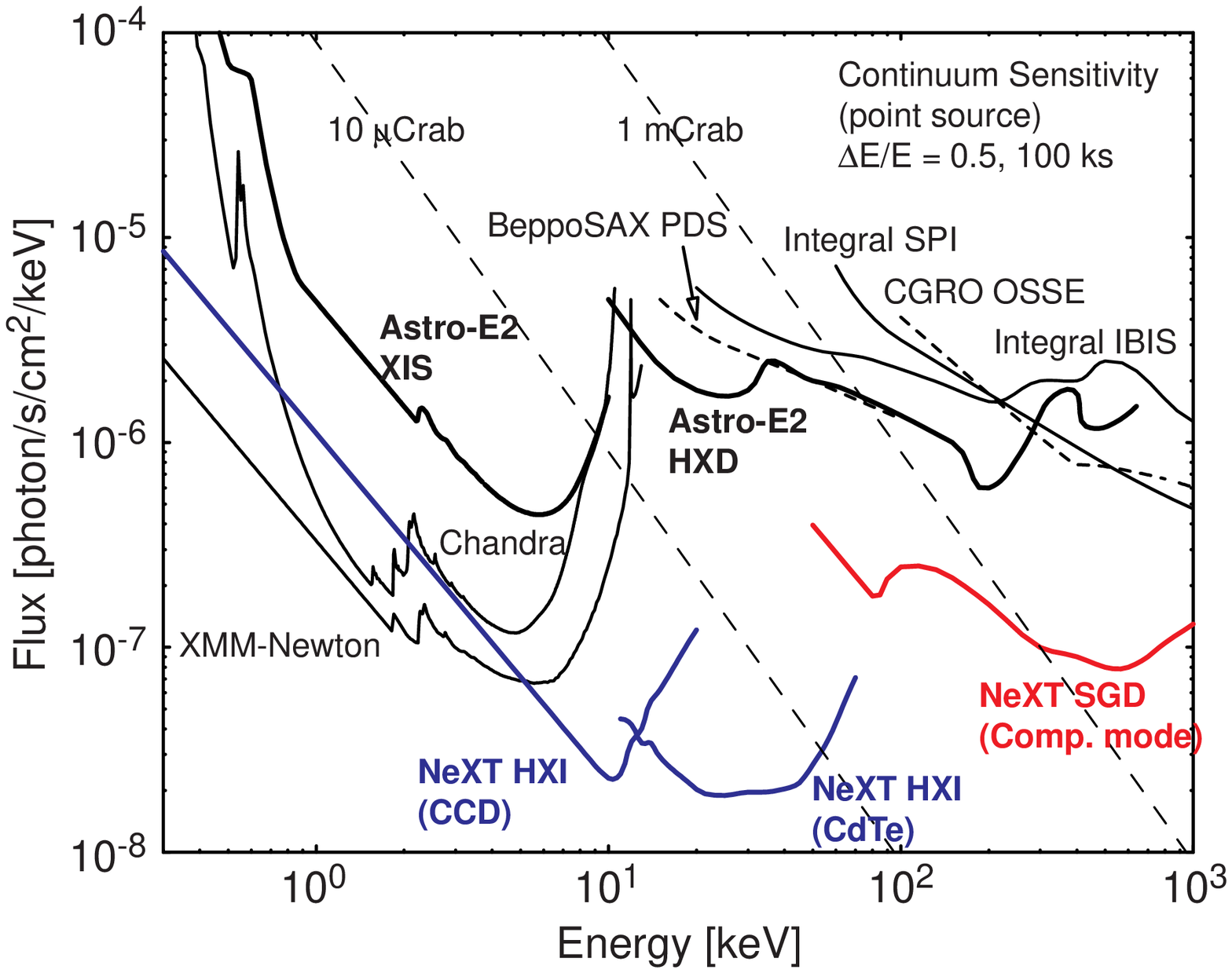}
   \caption[] 
   { \label{fig:NeXT-sensitivity} 3$\sigma$ sensitivity targets for the SXI, HXI and SGD in the NeXT mission for continuum emissions from point source, assuming an observation time of 100~ks and comparison with other instruments.}
   \end{figure}

The NeXT/SGD is a Compton telescope with narrow field of view, which utilizes Compton kinematics to enhance its background rejection capabilities.
It is realized as a hybrid semiconductor gamma-ray detector which consists of 
silicon and CdTe detectors.
Fig.~\ref{fig:NeXT-sensitivity} shows 3$\sigma$ sensitivity targets for three instruments in the NeXT mission, the Soft X-ray Imager (SXI), the Hard X-ray Imager(HXI) and SGD, for continuum emission from point source with an observation time of 100~ks\cite{Takahashi04-SGD} and comparison with other instruments. (Sensitivity depends on the bandwidth of each point and observation time, and can be lower than the background level with sufficient statistics.)
The SGD presents great improvement in the soft gamma-ray band compared with the currently operating INTEGRAL\cite{INTEGRAL} or Astro-E2 HXD to be launched in 2005,
and extends the bandpass to well above the cutoff for the hard X-ray telescope, which in turn allows us to study the high energy end of the particle spectrum.
Combined with the SXI and the HXI on board the NeXT, the SGD realizes unprecedented level of sensitivities from soft X-ray to soft gammma-ray band.
We also expect $5\sigma$ polarization sensitivity for 10~mCrab sources with 11\% intrinsic polarization.

We have developed low noise analog ASICs and CdTe detector technologies required to realize the SGD.
We have achieved 1.1~keV and 1.3~keV energy resolution for silicon strip detectors at 14~keV and 60--120~keV, respectively, and 1.7~keV for CdTe pixel detectors at 60~keV.
We have verified that the APD readout of BGO crystal can detect the energy deposit down to 11~keV.

We have demonstrated the performance of prototype Si/CdTe Compton telescopes.
MC simulations are validated against the experimental data on the angular resolution and the polarization performance.
Doppler broadening, not the energy resolution or position resolution, is confirmed to be the dominant contribution.
These experimental results validate the SGD design and specifications.


\bibliographystyle{IEEEtran.bst}
\bibliography{mybib}   

\end{document}